\documentclass[superscriptaddress,nobibnotes,amsmath,amssymb,notitlepage,twocolumn,pra,longbibliography]{revtex4-2}

\usepackage{bm,braket}
\usepackage[toc,page]{appendix}
\usepackage{comment}
\usepackage{dcolumn}

\usepackage{amsfonts}

\usepackage{color,hyperref}
\usepackage[caption=false]{subfig}
\hypersetup{colorlinks=true, linkcolor=blue, citecolor=blue, urlcolor=blue} 

\newcommand{\beq}[1]{\begin{equation}\label{#1}}
\newcommand{\eep}{\;.\end{equation}}
\newcommand{\eec}{\;,\end{equation}}
\newcommand{\eeq}{\end{equation}}

\newcommand{\om}{\omega}

\DeclareMathAlphabet{\mathcal}{OMS}{cmsy}{m}{n} 

\usepackage{amsmath}
\usepackage{amssymb}
\usepackage{xcolor}
\usepackage{bbm}
\usepackage{physics}
\usepackage{float}
\usepackage{dcolumn} 
\usepackage{bm} 
\usepackage{siunitx}
\usepackage{enumitem}  

\makeatletter
\renewcommand*{\fnum@figure}{{\normalfont\bfseries \figurename~\thefigure}}
\makeatother

\allowdisplaybreaks

\definecolor{orange}{rgb}{1,0.5,0}

\newcommand{\sect}[1]{\vspace{0.3em}{\it #1.}---}

\DeclareMathAlphabet{\mathcal}{OMS}{cmsy}{m}{n} 

\makeatletter
\newcommand{\specificthanks}[1]{\@fnsymbol{#1}}
\makeatother

\begin{document}

\preprint{APS/123-QED}

\title{Multistate Torsion in \textit{Zitterbewegung}}

\author{Wojciech J. Jankowski}
\email{wjj25@cam.ac.uk}
\affiliation{TCM Group, Cavendish Laboratory, Department of Physics, J J Thomson Avenue, Cambridge CB3 0HE, United Kingdom}

\date{\today}

\begin{abstract}

We derive nonlinear \textit{Zitterbewegung} velocity of electrons augmented by band torsion in crystals. The equation of motion with multistate quantum geometric tensors and torsion culminates in transient nonlinear photocurrent responses distinct from the recently measured resonant steady-state photovoltaic effects induced by circularly~polarized light. Tomography protocols for measuring the multistate quantum geometric torsion tensors are proposed.
     
\end{abstract} 

\maketitle

\sect{Introduction} Quantum geometry in condensed \mbox{matter} physics provides a versatile paradigm for studying multiband effects in solids~\cite{Torma2023, Bouhon2023, Yu2025QG, Verma2026, Gao2026}. Applications of the quantum geometric frameworks range from superconductivity~\cite{Peotta2015, Xie2020, Arbeitman2022}, magnetism~\cite{Piechon2016, Chau2026, Yuan2026, Hu2026}, or excitonics~\cite{Yao2008, Jankowski2025exciton, Thompson2025exciton, Davenport2026, Choi2026}, to~linear and nonlinear transport and optics~\cite{Nagaosa2010, Gao2014, Sodemann2015, Ahn2020, Ahn2021}. In~spirit, these frameworks can be systematically reduced to identifying gauge-invariant quantum tensors defined by inner products of quantum state projectors and their derivatives~\cite{Ahn2021, Mitscherling2025} within the perturbative structure of quantum response functions. This approach, underpinned by identification of quantum geometric bounds~\cite{Peotta2015, Xie2020, Arbeitman2022, Onishi2024, Kwon2024, Jankowski2024PRBHopf, Jankowski2025PRBoptical, Jankowski2025PRR, Jankowski2025PRL, Jankowski2025exciton, Shinada2025, Hung2026}, has proven particularly fruitful for studying physical consequences of nontrivial band topologies in solids, through their quantum geometric fingerprints. 

At heart of the quantum geometric frameworks is the study of the quantum geometric tensors (QGTs)~\cite{Provost1980,Marzari1997,Resta2011}, also known as Hermitian metrics of quantum states~\cite{Ahn2021, Gao2026}. As complex quantities, QGTs consist of a real part, the quantum metric~\cite{Provost1980}, and imaginary part, the Berry curvature equal to a geometric Berry phase over an infinitesimal loop in parameter space~\cite{Xiao2010, Vanderbilt2018}. The parameter space of choice in condensed matter systems is provided by crystal momenta $\textbf{k}$ of Bloch states over the Brillouin zone~(BZ). As gauge-invariant objects, the QGTs resolved over the momentum space have been measured in crystals only recently~\cite{Kang2025, Kim2025}. 

The nontrivial QGTs definitionally require a single quantum state, and a Hilbert space with at least two states~\cite{Provost1980}. However, it is further possible to define gauge-invariant two-state~\cite{Ahn2021, Mitscherling2025} and multistate QGTs beyond Berry curvature and quantum metric, such as band torsion~\cite{Ahn2020, Ahn2021, Mehraeen2025, Farrag2026}, which arise in nonlinear optics~\cite{Ahn2020, Ahn2021, Jankowski2024PRL, Avdoshkin2025, Guo2025} and transport phenomena~\cite{Mehraeen2025, Farrag2026} involving multiple energy levels. The interest in multistate quantum geometries has been particularly reinvigorated~\cite{Avdoshkin2025, Mehraeen2025, Bradlyn2026} with the discoveries of exotic nonlinear photovoltaic and viscoelastic effects involving higher-rank QGTs and band torsion~\cite{Jankowski2024PRL, Jain2026PRL, Jain2026TAD}. Notably, photovoltaic circular shift photoconductivities associated with band torsion have been experimentally measured in MnPSe$_3$ antiferromagnetic crystals very recently~\cite{Tian2026}.

On a seemingly different end, QGTs have been most recently revisited from the perspective of classical Hamiltonian dynamics in presence of effective magnetic fields~\cite{Stern2026}, subject to semiclassical quantizations of angular momentum. Yet, this opens up a~question about the interpretations and the role of multistate QGTs and band torsion in semiclassical dynamics~\cite{Ahn2021, Jankowski2024PRL, Jankowski2024PRBHopf, Mitscherling2025, Jankowski2025gerbe}. 

In this work, we address this question from an angle of the~semiclassical equations of motion for electron velocity in clean crystals. In doing so, we surprisingly find uncharted transient effects driven by band torsion. First, we demonstrate that the multistate QGTs and band torsion arise in the nonlinear $\textit{Zitterbewegung}$ velocity associated with nonequilibrium jittery trembling motion of~electrons under external electric fields. Second, we show that these multimode $\textit{Zitterbewegung}$ velocities of electrons culminate in transient nonlinear photocurrents due to the multistate torsion tensor. These photocurrents may be measured in an instantaneous nonlinear response to circularly~polarized optical fields. Finally, we argue that the multimode \textit{Zitterbewegung} enables a dynamical tomography of band torsion, allowing one to access this quantum geometric quantity directly, in contrast with the resonant response measurements. These findings apply to transient bulk responses of ultraclean crystalline materials and are further adaptable for ultracold atom quantum simulators.

\sect{Multiband Zitterbewegung} Before \mbox{deriving} the nonlinear multiband \textit{Zitterbewegung} central to this work, we~begin with the field-induced first-order \textit{Zitterbewegung} contributed by an interband coherence associated with the two-state quantum geometry. At first order in electric fields $E_j$, the equation for an electron velocity $v^m_i (t)$ in a nondegenerate band with index $m$ in a clean crystal reads \cite{Culcer2026},
\beq{}
   v^m_i (t)  =\frac{1}{\hbar} \partial_i \varepsilon_m + \frac{e E_j }{\hbar} \Omega^m_{ij} + \xi^{m}_i(t).
\eeq
The momentum-space derivative, $\partial_i \equiv \partial_{k_i}$, with ${i,j=x,y,z}$, acts on the band energy $\varepsilon_m$ in a dispersive drift velocity term, and Berry curvature $\Omega^m_{ij}$ contributes a~linear Hall drift of an electron. The last term represents the static-field-induced first-order \textit{Zitterbewegung} velocity~\cite{Culcer2026},
\beq{}\label{eq::xi1}
    \xi^{m,(1)}_i(t) = \frac{\text{i} e E_j}{\hbar} \sum_{n \neq m} Q^{nm}_{ij} e^{-\text{i} \om_{mn} t} + \text{h.c.},
\eeq
with frequency $\om_{mn} = (\varepsilon_m - \varepsilon_n)/\hbar$, two-state QGT, $Q^{nm}_{ij} = A^{nm}_i A^{mn}_j$, and non-Abelian Berry connection, $A^{nm}_i = \text{i} \bra{u_{n\textbf{k}}}\ket{\partial_{i}u_{m\textbf{k}}}$, derived from Bloch states $\ket{u_{m\textbf{k}}}$. Upon promoting the static electric field $E_j$ to an optical field $E_j(t) = \text{Re}[E_j(\om) e^{\text{i} \om t}]$, a dynamical field-induced first-order \textit{Zitterbewegung} velocity arises, as we show in the Supplemental Material (SM)~\cite{SM}. The dynamically-induced \textit{Zitterbewegung} provides a microscopic physical interpretation of the charge dipole correlations in the framework of time-dependent quantum geometry~\cite{Verma2025PNAS}.

Centrally to this work, at second order in the optical electric fields~$E_j(\om)$, a nonlinear multimode \textit{Zitterbewegung} velocity further arises,
\beq{}\label{eq::xi2}
    \xi^{m,(2)}_i(t) = \frac{ e^2 E_j(\om) E_k (\eta \om)}{\hbar^2} \sum_{a,b = n,p,m}  \mathcal{A}^{ab}_{ijk}(\omega, \eta \omega) e^{-\text{i} \om_{ab} t},
\eeq
with $\eta = \pm$ corresponding to distinct frequency combination channels. The frequency-dependent factors $\mathcal{A}^{ab}_{ijk}(\omega, \eta \omega)$ encode the two-state and three-state QGTs, $Q^{nmp}_{ijk} = A^{nm}_i A^{mp}_j A^{pn}_k$, as derived in the SM~\cite{SM}. Interestingly, a~nonlinear multimode \textit{Zitterbewegung} velocity in band $m$ includes modes with frequencies $\om_{pn}$ such that $p,n \neq m$ correspond to a distinct pair of virtual or unoccupied \mbox{levels}. As a key result, we demonstrate that the multistate band torsion, $T^{nmp}_{ijk} = \text{i}(Q^{nmp}_{ijk} - Q^{nmp}_{ikj})$, arises in circular \textit{Zitterbewegung} contributed by modes with frequencies $\om_{pn}$ independent of the occupied band~$m$. As~a~consequence, the band torsion is directly manifested in the high-frequency modes, with $\om_{pn} > \om_{mn}, \om_{mp}$, of~transient photocurrents.

\sect{Nonlinear Zitterbewegung currents} Having derived the second-order \textit{Zitterbewegung} velocities, we focus on the associated transient photocurrents, which can be \mbox{induced} by an instantaneous application of optical fields. 

The bulk electric current $j_i$ can be computed from density matrix $\hat{\rho}$ and velocity operator $\hat{v}_i$,
\beq{}
    j_i (t) = -e \text{Tr} [\hat{\rho} \hat{v}_i] = -e \sum_{n,m, \textbf{k}} \rho_{nm} v^{mn}_i.
\eeq
For $n \neq m$, $v^{nm}_i = \text{i} \om_{nm} A^{nm}_i$, and the single-particle density matrix interband coherences $\rho_{nm}$ can be perturbatively expanded as ${\rho_{nm} =\rho^{(0)}_{nm} +  \rho^{(1)}_{nm} + \rho^{(2)}_{nm} + \ldots}$, while satisfying quantum Liouville equation~\cite{Mandal2024, Culcer2026}. Centrally to this work, we consider the interband coherences up to the second order, $\rho^{(2)}_{nm}$, in the optical fields. The multistate band torsion arises in the three-state contributions to $\rho^{(2)}_{nm}$, which translate into the nonlinear \textit{Zitterbewegung} velocity, $\xi^{m,(2)}_i(t)$, as demonstrated in the SM~\cite{SM}.

The second-order field-induced \textit{Zitterbewegung} currents which arise in transient nonlinear responses read, ${j^{(2)}_i(t) = -e \sum_m f_m \xi^{m,(2)}_i(t)}$. Centrally to this work, these currents include multistate band torsion $T^{mnp}_{ijk}$ contributions in instantaneous responses to electric fields with magnitude $E_0$,
\beq{}
    j^{(2)}_i (t) \Big|_{\om_{pn}} = -\frac{e^3 E^2_0}{\hbar^2} \sum_{m,n,p,\textbf{k}} f_m w^{+-}_{nmp}(\om) \text{Im}~\Big[ T^{nmp}_{ijk} e^{-\text{i} \om_{pn} t} \Big].
\eeq
with the off-resonant frequency-dependent factors $w^{+-}_{nmp}(\om) =  \frac{\om (\om^2_{pm} - \om^2_{mn})}{(\om^2_{mn} - \om^2)(\om^2_{pm} - \om^2)}$. In the above, we consider the counter-propagating channel, $\eta = -$, and a response to the left circularly~polarized light with an electric field vector, $\textbf{E}(t) =  \Theta (t) E_0[\hat{x}~\text{cos}(\om t) + \hat{y}~\text{sin}(\om t)]$, $\Theta(t)$ the unit Heaviside step function, and $\hat{x}$, $\hat{y}$ the unit vectors in the $x$ and $y$ directions. For the derivation of the corresponding \textit{Zitterbewegung}, see SM~\cite{SM}. Notably, under the linear polarization, $\textbf{E}(t) =  \Theta (t) E_0[\hat{x}~\text{cos}(\om t)]$,
\beq{}
    j^{(2)}_i (t) \Big|_{\om_{pn}} = -\frac{e^3 E^2_0}{\hbar^2} \sum_{m,n,p,\textbf{k}} f_m w^{\hat{x} \hat{x}}_{mnp}(\om) \text{Im}~\Big[ Q^{nmp}_{ixx} e^{-\text{i} \om_{pn} t} \Big],
\eeq
with the distinct off-resonant frequency factors, ${w^{\hat{x} \hat{x}}_{mnp}(\om) = \sum_{\eta = \pm} \frac{\om_{pn} - \eta \om}{(\om_{pm} - \eta \om)(\om_{mn} - \eta \om)}}$. Therefore, the nonlinear \textit{Zitterbewegung} currents within a second-order transient response to linearly polarized optical electric fields allows to probe the three-state QGTs $Q^{nmp}_{iaa}$ with $a= x,y$. Thus, we have shown that controlling the polarization of optical fields allows to determine distinct multistate geometric tensors.

%
%

\sect{Discussion and conclusion} We further discuss the~applicability and significance of derived results.

First, we establish the second-order circular \textit{Zitterbewegung} velocities and the associated transient electric currents as the direct smoking-gun probes of band torsion. It~should be stressed that, contrary to the derived circular \textit{Zitterbewegung}, the band torsion, ${T^{mn}_{ijk} = \sum_{p\neq n,m} T^{mnp}_{ijk}}$, does not arise in the individual resonant ($\om = \om_{mn}$) circular shift photoconductivity tensor components \textit{per se}. In fact, the band torsion contributes only the totally antisymmetric part of the circular shift photoconductivity tensor, ${\mathcal{H}^{nm}_{xyz} \equiv T^{nm}_{[xyz]}}$~\cite{Jankowski2024PRL, Bzdusek2026, Jankowski2025gerbe}, where $[\ldots]$ denotes an antisymmetrization in all indices. Therefore, the band torsion cannot be directly reconstructed from the single-component measurements of resonant circular shift currents, while the momentum space average of its totally antisymmetric part can only be extracted after measuring multiple components of the shift photoconductivity tensor $(\sigma^{\text{shift,C}}_{xyz}, \sigma^{\text{shift,C}}_{yzx}, \sigma^{\text{shift,C}}_{zxy})$~\cite{Jankowski2024PRL}. It should be stressed that, on the contrary, the retrieved \textit{Zitterbewegung} currents under circular drive allow to resolve individual torsion tensor components directly. Therefore, this work establishes the second-order \textit{Zitterbewegung} currents as the first-of-a-kind quantity allowing to directly access the quantum geometric multistate band torsion components, without requiring resonant drive.

Second, we compare the retrieved nonlinear multiband \textit{Zitterbewegung} with the other known quantum trembling motions of electrons in crystals. In the absence of~electric fields, the \textit{Zitterbewegung} amplitudes $Z_{nm}$ determined by the interband coherences and captured by non-Abelian Berry connections $A^{nm}_i$~\cite{Gyula2010, Jozsef2010}, arise in the off-diagonal elements of position operators expressed in the band basis. Hence, the \textit{Zitterbewegung} amplitudes $Z_{nm}$ yield vanishing expectation values upon projecting the electron onto a pure band $m$. At first order in electric fields, the \textit{Zitterbewegung} velocity magnitude $|\xi^{m,(1)}_i|$ is proportional to the two-state QGTs, $Q^{nm}_{ij} = A^{nm}_i A^{nm}_j \propto Z^2_{nm}$, consistently with Eq.~\eqref{eq::xi1}~\cite{Jozsef2010, Culcer2026}. The interpretation of QGT as field-induced \textit{Zitterbewegung} provides a phenomenological picture for the recent proposal of measuring quantum metric from an instantaneous step response~\cite{Verma2025, Culcer2026}. At second order in electric fields, centrally to this work, ${Q^{nmp}_{ijk} = A^{nm}_i A^{mp}_j A^{pn}_k \propto Z_{nm} Z_{mp} Z_{pn}}$, which provides a, ~cubic in \textit{Zitterbewegung} amplitudes, interpretation to the second-order field-induced \textit{Zitterbewegung} velocity magnitude $|\xi^{m,(2)}_i|$ in Eq.~\eqref{eq::xi2}, with ${\mathcal{A}^{np}_{ijk}(\omega, \eta \omega) \propto Q^{nmp}_{ijk}}$. On the other hand, the multistate torsion ${T^{nmp}_{ijk} \propto Q^{nmp}_{ijk} - Q^{nmp}_{ikj}}$ is represented by the antisymmetrization of \textit{Zitterbewegung} amplitudes picked by the circularly~polarized light coupling. At~third order in electric fields, we expect a four-state QGT, ${Q^{nmpq}_{ijkl} = A^{nm}_i A^{mp}_j A^{pq}_k A^{qn}_l \propto Z_{nm} Z_{mp} Z_{pq} Z_{qn}}$, which is quartic in \textit{Zitterbewegung} amplitudes, to determine the field-induced third-order \textit{Zitterbewegung} velocity~$|\xi^{m,(3)}_i|$. Phenomenologically, $\xi^{m,(3)}_i$ underpins the quantum \mbox{geometric} enhancement of hyperpolarizability~$\chi^{(3)}$~\cite{Jankowski2025PRL}. Hence, the discussed findings establish a hierarchy of \mbox{\textit{Zitterbewegung}} amplitude scalings in nonlinear quantum geometric responses.

Third, we discuss the measurability of the nonlinear \textit{Zitterbewegung} effects driven by the multistate \mbox{torsion}. Although generally underexplored in real materials, the~multimode \textit{Zitterbewegung} of electrons has been previously considered in graphene~\cite{Rusin2013}. We expect the derived transient responses to be highly dependent on the disorder present in material, with the framework of this work being applicable only in the ultraclean system limit with extremely high scattering times~$\tau$. Nevertheless, we~expect the derived intrinsic transient \mbox{nonlinear} \mbox{\textit{Zitterbewegung}} photocurrents to persist  for any finite scattering times $\tau$. The study of the effects of \mbox{disorder} and interactions on the quantum geometric nonlinear \mbox{\textit{Zitterbewegung}} currents is left for future work.

Finally, we propose a dynamical protocol for extracting multistate torsion tensor and QGTs from transient current tomography. To reconstruct the multistate torsion, the Fourier components of the current at virtual band energy differences, $\om_{pn}$, are of particular interest. Different drive frequencies $\omega$ and light polarizations allow to determine and control the amplitude factors $w^{+-}_{nmp}(\om)$, $w^{\hat{x} \hat{x}}_{nmp}(\om)$, beyond the filling factors $f_m$ controllable with temperature and chemical potential. Acquisition of the dynamics of the Fourier modes with frequencies $\om_{pn}(\textbf{k})$ in the transient photocurrents thus allows to estimate $T^{nmp}_{ijk}$ and $Q^{nmp}_{ijj}$ averaged over a constant frequency $(\om_{pn})$ contour  determined by the band energy splittings.

It should be noted that the zero-field (free) \textit{Zitterbewegung} proportional to amplitudes $Z_{ab}$ has been measured in ultracold atom quantum simulators~\cite{Vaishnav2008}. More \mbox{recently}, a~protocol to reconstruct the interband Berry connections that implicitly probes the \textit{Zitterbewegung} amplitudes, ${A^{nm}_i \propto Z_{nm}}$, has been successfully developed in optical lattices~\cite{Chang2026}. In the latter, the shaking force $F$ on optical lattice acts as an effective electric field that couples to the position operator, analogously to the~Hamiltonian perturbation in the quantum Liouville equations. As a consequence, the nonlinear correction to the density matrix, $\rho^{(2)}_{nm} \propto F^2$, can be analogously induced, and its fingerprint in the nonlinear \textit{Zitterbewegung} velocity $\xi^{m,(2)}_i(t)$ may be similarly measured.

To sum up, we elucidated the role of band torsion in the generalized equations of motion for the electron velocities in multiband systems. The multistate torsion and QGTs were shown to arise in the multimode \textit{Zitterbewegung} induced by the second-order coupling to optical electric fields. The retrieved \textit{Zitterbewegung} is manifested in nonlinear transient photocurrents, providing a~direct off-resonant probe of multistate band torsion and QGTs.
    
\sect{Acknowledgements} W.J.J.~acknowledges funding from the Rod Smallwood Studentship at Trinity College, Cambridge. The author thanks G.~Chaudhary, A.~Jain, M.~Mehraeen, G.~Palumbo, and Anthropic's Claude, \mbox{Fable 5.1}, for helpful discussions.

\bibliography{references}

@misc{Tian2026,
      title={Observation of magnetically switchable quantum geometric photocurrents}, 
      author={Qi Tian and Zhuoliang Ni and Matthew Cothrine and David G. Mandrus and Eugene J. Mele and Andrew M. Rappe and Charles L. Kane and Fernando de Juan and Liang Wu},
      year={2026},
      eprint={2605.22518},
      archivePrefix={arXiv},
      primaryClass={cond-mat.mtrl-sci},
      url={https://arxiv.org/abs/2605.22518}, 
}

@misc{Stern2026,
      title={How quantum is quantum geometry?}, 
      author={Ady Stern and Felix von Oppen},
      year={2026},
      eprint={2608.26269},
      archivePrefix={arXiv},
      primaryClass={cond-mat.mes-hall},
      url={https://arxiv.org/abs/2608.26269}, 
}

@article{Torma2023,
  title = {Essay: Where Can Quantum Geometry Lead Us?},
  author = {T\"orm\"a, P\"aivi},
  journal = {Phys. Rev. Lett.},
  volume = {131},
  issue = {24},
  pages = {240001},
  numpages = {7},
  year = {2023},
  month = {Dec},
  publisher = {American Physical Society},
  doi = {10.1103/PhysRevLett.131.240001},
  url = {https://link.aps.org/doi/10.1103/PhysRevLett.131.240001}
}

@article{Gao2026,
  title = {Quantum geometry phenomena in condensed matter systems},
  author = {Gao, Anyuan and Nagaosa, Naoto and Ni, Ni and Xu, Su-Yang},
  journal = {Rev. Mod. Phys.},
  volume = {98},
  issue = {3},
  pages = {035005},
  numpages = {67},
  year = {2026},
  month = {Sep},
  publisher = {American Physical Society},
  doi = {10.1103/f78t-ky69},
  url = {https://link.aps.org/doi/10.1103/f78t-ky69}
}

@Article{Jankowski2025exciton,
  author = {Jankowski, W. J.
and Thompson, J. J. P.
and Monserrat, B.
and Slager, R.-J.},
  day = {19},
  issn = {2041-1723},
  journal = {Nat. Commun.},
  number = {1},
  pages = {4661},
  title = {Excitonic topology and quantum geometry in organic semiconductors},
  url = {https://doi.org/10.1038/s41467-025-59257-5},
  volume = {16},
  year = {2025}
}

@article{Peotta2015,
  author = {Peotta, S. and T{\"o}rm{\"a}, P.},
  journal = {Nat. Commun.},
  number = {1},
  pages = {1--9},
  publisher = {Nature Publishing Group},
  title = {Superfluidity in topologically nontrivial flat bands},
  url = {https://www.nature.com/articles/ncomms9944},
  volume = {6},
  year = {2015}
}

@article{Xie2020,
  author = {Xie, F. and Song, Z. and Lian, B. and Bernevig, B. A.},
  issue = {16},
  journal = {Phys. Rev. Lett.},
  numpages = {6},
  pages = {167002},
  publisher = {American Physical Society},
  title = {Topology-Bounded Superfluid Weight in Twisted Bilayer Graphene},
  url = {https://link.aps.org/doi/10.1103/PhysRevLett.124.167002},
  volume = {124},
  year = {2020}
}

@article{Arbeitman2022,
  title = {Superfluid Weight Bounds from Symmetry and Quantum Geometry in Flat Bands},
  author = {Herzog-Arbeitman, Jonah and Peri, Valerio and Schindler, Frank and Huber, Sebastian D. and Bernevig, B. Andrei},
  journal = {Phys. Rev. Lett.},
  volume = {128},
  issue = {8},
  pages = {087002},
  numpages = {8},
  year = {2022},
  month = {Feb},
  publisher = {American Physical Society},
  doi = {10.1103/PhysRevLett.128.087002},
  url = {https://link.aps.org/doi/10.1103/PhysRevLett.128.087002}
}

@article{Ahn2021,
  author = {Ahn, J. and Guo, G.-Y. and Nagaosa, N. and Vishwanath, A.},
  issn = {1745-2481},
  journal = {Nat. Phys.},
  number = {3},
  pages = {290–295},
  publisher = {Springer Science and Business Media LLC},
  title = {Riemannian geometry of resonant optical responses},
  url = {http://dx.doi.org/10.1038/s41567-021-01465-z},
  volume = {18},
  year = {2022}
}

@article{Ahn2020,
  title = {Low-Frequency Divergence and Quantum Geometry of the Bulk Photovoltaic Effect in Topological Semimetals},
  author = {Ahn, Junyeong and Guo, Guang-Yu and Nagaosa, Naoto},
  journal = {Phys. Rev. X},
  volume = {10},
  issue = {4},
  pages = {041041},
  numpages = {28},
  year = {2020},
  month = {Nov},
  publisher = {American Physical Society},
  doi = {10.1103/PhysRevX.10.041041},
  url = {https://link.aps.org/doi/10.1103/PhysRevX.10.041041}
}

@article{Nagaosa2010,
  author    = {Nagaosa, Naoto and Sinova, Jairo and Onoda, Shigeki
               and MacDonald, A. H. and Ong, N. P.},
  title     = {{Anomalous Hall effect}},
  journal   = {Rev. Mod. Phys.},
  volume    = {82},
  number    = {2},
  pages     = {1539--1592},
  year      = {2010},
  month     = {May},
  doi       = {10.1103/RevModPhys.82.1539}
}

@article{Gao2014,
  author    = {Gao, Yang and Yang, Shengyuan A. and Niu, Qian},
  title     = {{Field Induced Positional Shift of Bloch Electrons
               and Its Dynamical Implications}},
  journal   = {Phys. Rev. Lett.},
  volume    = {112},
  number    = {16},
  pages     = {166601},
  year      = {2014},
  month     = apr,
  doi       = {10.1103/PhysRevLett.112.166601}
}

@article{Sodemann2015,
  author    = {Sodemann, Inti and Fu, Liang},
  title     = {{Quantum nonlinear Hall effect induced by Berry curvature dipole
               in time-reversal invariant materials}},
  journal   = {Phys. Rev. Lett.},
  volume    = {115},
  number    = {21},
  pages     = {216806},
  year      = {2015},
  month     = nov,
  doi       = {10.1103/PhysRevLett.115.216806}
}

@misc{SM,
note = {See Supplemental Material ({SM}) for further details on the multistate quantum geometry ({S}ec.~{I}), and the derivations of dynamically-induced Zitterbewegungs at first order ({S}ec.~{II}) and second order ({S}ec.~{III}) in optical electric fields.}
}

@article{Rusin2013,
  title = {{Multimode behavior of electron Zitterbewegung induced by an electromagnetic wave in graphene}},
  author = {Rusin, Tomasz M. and Zawadzki, Wlodek},
  journal = {Phys. Rev. B},
  volume = {88},
  issue = {23},
  pages = {235404},
  numpages = {15},
  year = {2013},
  month = {Dec},
  publisher = {American Physical Society},
  doi = {10.1103/PhysRevB.88.235404},
  url = {https://link.aps.org/doi/10.1103/PhysRevB.88.235404}
}

@article{Mehraeen2025,
  title = {Quantum Response Theory and Momentum-Space Gravity},
  author = {Mehraeen, M.},
  journal = {Phys. Rev. Lett.},
  volume = {135},
  issue = {15},
  pages = {156302},
  numpages = {7},
  year = {2025},
  month = {Oct},
  publisher = {American Physical Society},
  doi = {10.1103/t6nt-qzws},
  url = {https://link.aps.org/doi/10.1103/t6nt-qzws}
}

@misc{Farrag2026,
      title={{High Order Geometric Channels for Nonlinear Transport in Bloch Bands}}, 
      author={Sami Farrag and Eugene Mele and Tony Low},
      year={2026},
      eprint={2607.20702},
      archivePrefix={arXiv},
      primaryClass={cond-mat.mes-hall},
      url={https://arxiv.org/abs/2607.20702}, 
}

@article{Gyula2010,
  title = {{General theory of Zitterbewegung}},
  author = {D\'avid, Gyula and Cserti, J\'ozsef},
  journal = {Phys. Rev. B},
  volume = {81},
  issue = {12},
  pages = {121417(R)},
  numpages = {4},
  year = {2010},
  month = {Mar},
  publisher = {American Physical Society},
  doi = {10.1103/PhysRevB.81.121417},
  url = {https://link.aps.org/doi/10.1103/PhysRevB.81.121417}
}

@article{Jozsef2010,
  title = {{Relation between Zitterbewegung and the charge conductivity, Berry curvature, and the Chern number of multiband systems}},
  author = {{Cserti, J\'ozsef and D\'avid, Gyula}},
  journal = {Phys. Rev. B},
  volume = {82},
  issue = {20},
  pages = {201405(R)},
  numpages = {4},
  year = {2010},
  month = {Nov},
  publisher = {American Physical Society},
  doi = {10.1103/PhysRevB.82.201405},
  url = {https://link.aps.org/doi/10.1103/PhysRevB.82.201405}
}

@article{Vaishnav2008,
  title = {{Observing Zitterbewegung with Ultracold Atoms}},
  author = {Vaishnav, J. Y. and Clark, Charles W.},
  journal = {Phys. Rev. Lett.},
  volume = {100},
  issue = {15},
  pages = {153002},
  numpages = {4},
  year = {2008},
  month = {Apr},
  publisher = {American Physical Society},
  doi = {10.1103/PhysRevLett.100.153002},
  url = {https://link.aps.org/doi/10.1103/PhysRevLett.100.153002}
}

@misc{Chang2026,
      title={{Interband Berry connection measurement in the optical honeycomb lattice}}, 
      author={Shao-Wen Chang and Malte N. Schwarz and Erin G. Moloney and Ke Lin and Dan M. Stamper-Kurn},
      year={2026},
      eprint={2605.11597},
      archivePrefix={arXiv},
      primaryClass={cond-mat.quant-gas},
      url={https://arxiv.org/abs/2605.11597}, 
}

@article{Guo2025,
  title = {Bicircular light-induced multistate geometric current},
  author = {Guo, Zhichao and Lu, Zhuocheng and Wang, Hua and Chang, Kai},
  journal = {Phys. Rev. B},
  volume = {112},
  issue = {3},
  pages = {035162},
  numpages = {24},
  year = {2025},
  month = {Jul},
  publisher = {American Physical Society},
  doi = {10.1103/7ytw-vyb7},
  url = {https://link.aps.org/doi/10.1103/7ytw-vyb7}
}

@article{Verma2025PNAS,
  author    = {Verma, Nishchhal and Queiroz, Raquel},
  title     = {Instantaneous response and quantum geometry of insulators},
  journal   = {Proceedings of the National Academy of Sciences},
  year      = {2025},
  volume    = {122},
  number    = {49},
  pages     = {e2405837122},
  doi       = {10.1073/pnas.2405837122},
  issn      = {0027-8424},
  publisher = {National Academy of Sciences},
  month     = dec
}

@article{Jain2026PRL,
  author  = {Jain, Ashwat and Jankowski, Wojciech J. and Mehraeen, M. and Slager, Robert-Jan},
  title   = {{Nonlinear Odd Viscoelastic Effect}},
  journal = {Phys. Rev. Lett.},
  volume  = {137},
  number  = {3},
  pages   = {036302},
  year    = {2026},
  doi     = {10.1103/jg6l-gzfr}
}

@misc{Jain2026TAD,
      title={Topological Acoustic Diode}, 
      author={Ashwat Jain and Wojciech J. Jankowski and M. Mehraeen and Robert-Jan Slager},
      year={2026},
      eprint={2601.20951},
      archivePrefix={arXiv},
      primaryClass={cond-mat.mes-hall},
      url={https://arxiv.org/abs/2601.20951}, 
}

@article{Yuan2026,
  title = {{Quantum Geometry of Altermagnetic Magnons Probed by Light}},
  author = {Yuan, Rundong and Jankowski, Wojciech J. and Shen, Ka and Slager, Robert-Jan},
  journal = {Phys. Rev. Lett.},
  volume = {137},
  issue = {10},
  pages = {106901},
  numpages = {10},
  year = {2026},
  month = {Sep},
  publisher = {American Physical Society},
  doi = {10.1103/12cl-b9jj},
  url = {https://link.aps.org/doi/10.1103/12cl-b9jj}
}

@article{Avdoshkin2025,
  author  = {Avdoshkin, Alexander and Mitscherling, Johannes and Moore, Joel E.},
  title   = {Multistate Geometry of Shift Current and Polarization},
  journal = {Phys. Rev. Lett.},
  volume  = {135},
  number  = {6},
  pages   = {066901},
  year    = {2025},
  doi     = {10.1103/w761-8nf7}
}

@article{Kwon2024,
  title = {{Quantum geometric bound and ideal condition for Euler band topology}},
  author = {Kwon, Soonhyun and Yang, Bohm-Jung},
  journal = {Phys. Rev. B},
  volume = {109},
  issue = {16},
  pages = {L161111},
  numpages = {7},
  year = {2024},
  month = {Apr},
  publisher = {American Physical Society},
  doi = {10.1103/PhysRevB.109.L161111},
  url = {https://link.aps.org/doi/10.1103/PhysRevB.109.L161111}
}

@article{Verma2025,
  title = {Framework to Measure Quantum Metric from Step Response},
  author = {Verma, Nishchhal and Queiroz, Raquel},
  journal = {Phys. Rev. Lett.},
  volume = {134},
  issue = {10},
  pages = {106403},
  numpages = {6},
  year = {2025},
  month = {Mar},
  publisher = {American Physical Society},
  doi = {10.1103/PhysRevLett.134.106403},
  url = {https://link.aps.org/doi/10.1103/PhysRevLett.134.106403}
}

@article{Mandal2024,
  title = {Quantum geometry induced third-order nonlinear transport responses},
  author = {Mandal, Debottam and Sarkar, Sanjay and Das, Kamal and Agarwal, Amit},
  journal = {Phys. Rev. B},
  volume = {110},
  issue = {19},
  pages = {195131},
  numpages = {18},
  year = {2024},
  month = {Nov},
  publisher = {American Physical Society},
  doi = {10.1103/PhysRevB.110.195131},
  url = {https://link.aps.org/doi/10.1103/PhysRevB.110.195131}
}

@article{Xiao2010,
  title = {Berry phase effects on electronic properties},
  author = {Xiao, Di and Chang, Ming-Che and Niu, Qian},
  journal = {Rev. Mod. Phys.},
  volume = {82},
  issue = {3},
  pages = {1959--2007},
  numpages = {0},
  year = {2010},
  month = {Jul},
  publisher = {American Physical Society},
  doi = {10.1103/RevModPhys.82.1959},
  url = {https://link.aps.org/doi/10.1103/RevModPhys.82.1959}
}

@book{Vanderbilt2018,
  author = {Vanderbilt, D.},
  publisher = {Cambridge University Press},
  title = {Berry phases in electronic structure theory: electric polarization, orbital magnetization and topological insulators},
  url = {https://doi.org/10.1017/9781316662205},
  year = {2018}
}

@article{Culcer2026,
  title = {Zitterbewegung velocity in semiclassical electron dynamics},
  author = {Culcer, Dimitrie},
  journal = {Phys. Rev. B},
  volume = {114},
  issue = {2},
  pages = {L020305},
  numpages = {6},
  year = {2026},
  month = {Jul},
  publisher = {American Physical Society},
  doi = {10.1103/r2rv-tb7f},
  url = {https://link.aps.org/doi/10.1103/r2rv-tb7f}
}

@article{Kim2025,
author = {Sunje Kim  and Yoonah Chung  and Yuting Qian  and Soobin Park  and Chris Jozwiak  and Eli Rotenberg  and Aaron Bostwick  and Keun Su Kim  and Bohm-Jung Yang },
title = {Direct measurement of the quantum metric tensor in solids},
journal = {Science},
volume = {388},
number = {6751},
pages = {1050-1054},
year = {2025},
doi = {10.1126/science.ado6049},
URL = {https://www.science.org/doi/abs/10.1126/science.ado6049},
eprint = {}}

@Article{Kang2025,
  author = {Kang, M.
and Kim, S.
and Qian, Y.
and Neves, P. M.
and Ye, L.
and Jung, J.
and Puntel, D.
and Mazzola, F.
and Fang, S.
and Jozwiak, C.
and Bostwick, A.
and Rotenberg, E.
and Fuji, J.
and Vobornik, I.
and Park, J.-H.
and Checkelsky, J. G.
and Yang, B.-J.
and Comin, R.},
  day = {01},
  issn = {1745-2481},
  journal = {Nat. Phys.},
  number = {1},
  pages = {110-117},
  title = {Measurements of the quantum geometric tensor in solids},
  url = {https://doi.org/10.1038/s41567-024-02678-8},
  volume = {21},
  year = {2025}
}

@article{Provost1980,
  author = {Provost, J. P. and Vallee, G.},
  journal = {Commun. Math. Phys.},
  number = {3},
  pages = {289--301},
  publisher = {Springer},
  title = {Riemannian structure on manifolds of quantum states},
  volume = {76},
  year = {1980}
,
  url = {https://doi.org/10.1007/BF02193559}
}

@misc{Bradlyn2026,
      title={Multi-State Geometry of Density Matrices and Rectification Sum Rules}, 
      author={Barry Bradlyn},
      year={2026},
      eprint={2608.06326},
      archivePrefix={arXiv},
      primaryClass={cond-mat.mes-hall},
      url={https://arxiv.org/abs/2608.06326}, 
}

@article{Jankowski2024PRBHopf,
  author = {Jankowski, W. J. and Morris, A. S. and Davoyan, Z. and Bouhon, A. and \"Unal, F. N. and Slager, R.-J.},
  issue = {7},
  journal = {Phys. Rev. B},
  numpages = {24},
  pages = {075135},
  publisher = {American Physical Society},
  title = {Non-{A}belian {H}opf-{E}uler insulators},
  url = {https://link.aps.org/doi/10.1103/PhysRevB.110.075135},
  volume = {110},
  year = {2024}
}

@misc{Davenport2026,
      title={Composite Quantum Geometry and Semiclassical Dynamics}, 
      author={Henry Davenport and Yoonseok Hwang and Johannes Knolle and Frank Schindler},
      year={2026},
      eprint={2606.12525},
      archivePrefix={arXiv},
      primaryClass={cond-mat.mes-hall},
      url={https://arxiv.org/abs/2606.12525}, 
}

@misc{Choi2026,
      title={Enhancement of exciton radius near a band-gap closing through quantum geometry}, 
      author={Jin-Hyung Choi and Sang-Hoon Han and Young-Kwon Han and Sun-Woo Kim and Jun-Won Rhim and Joshua J. P. Thompson},
      year={2026},
      eprint={2607.28731},
      archivePrefix={arXiv},
      primaryClass={cond-mat.mes-hall},
      url={https://arxiv.org/abs/2607.28731}, 
}

@article{Yao2008,
  author  = {Yao, Wang and Niu, Qian},
  title   = {{Berry Phase Effect on the Exciton Transport and on the Exciton Bose-Einstein Condensate}},
  journal = {Phys. Rev. Lett.},
  volume  = {101},
  pages   = {106401},
  year    = {2008},
  doi     = {10.1103/PhysRevLett.101.106401}
}

@article{Resta2011,
	Author = {Resta, R. },
	Da = {2011/01/01},
	Doi = {10.1140/epjb/e2010-10874-4},
	Id = {Resta2011},
	Isbn = {1434-6036},
	Journal = {The European Physical Journal B},
	Number = {2},
	Pages = {121--137},
	Title = {The insulating state of matter: a geometrical theory},
	Ty = {JOUR},
	Url = {https://doi.org/10.1140/epjb/e2010-10874-4},
	Volume = {79},
	Year = {2011}}

@article{Marzari1997,
  title = {Maximally localized generalized {W}annier functions for composite energy bands},
  author = {Marzari, Nicola and Vanderbilt, David},
  journal = {Phys. Rev. B},
  volume = {56},
  issue = {20},
  pages = {12847--12865},
  numpages = {0},
  year = {1997},
  month = {Nov},
  publisher = {American Physical Society},
  doi = {10.1103/PhysRevB.56.12847},
  url = {https://link.aps.org/doi/10.1103/PhysRevB.56.12847}
}

@article{Onishi2024,
  author = {Onishi, Y. and Fu, L.},
  issue = {1},
  journal = {Phys. Rev. X},
  numpages = {12},
  pages = {011052},
  publisher = {American Physical Society},
  title = {Fundamental Bound on Topological Gap},
  url = {https://link.aps.org/doi/10.1103/PhysRevX.14.011052},
  volume = {14},
  year = {2024}
}

@article{Hung2026,
  title = {Extending Topological Bound on Quantum Weight beyond Symmetry-Protected Topological Phases},
  author = {Hung, Yi-Chun and Onishi, Yugo and Lin, Hsin and Fu, Liang and Bansil, Arun},
  journal = {Phys. Rev. Lett.},
  volume = {137},
  issue = {12},
  pages = {126601},
  numpages = {7},
  year = {2026},
  month = {Sep},
  publisher = {American Physical Society},
  doi = {10.1103/vmhd-jn5y},
  url = {https://link.aps.org/doi/10.1103/vmhd-jn5y}
}

@article{Shinada2025,
  author  = {Shinada, Koki and Nagaosa, Naoto},
  title   = {Quantum geometric bounds for observables: Linear responses, Drude weight, and orbital magnetization},
  journal = {Phys. Rev. B},
  volume  = {112},
  pages   = {155158},
  year    = {2025},
  doi     = {10.1103/qxbl-qd4f}
}

@article{Jankowski2025PRBoptical,
  author = {Jankowski, W. J. and Morris, A. S. and Bouhon, A. and \"Unal, F. N. and Slager, R.-J.},
  issue = {8},
  journal = {Phys. Rev. B},
  numpages = {7},
  pages = {L081103},
  publisher = {American Physical Society},
  title = {Optical manifestations and bounds of topological {E}uler class},
  url = {https://link.aps.org/doi/10.1103/PhysRevB.111.L081103},
  volume = {111},
  year = {2025}
}

@article{Jankowski2025PRR,
  author = {Jankowski, W. J. and Slager, R.-J. and Lange, G. F.},
  issue = {4},
  journal = {Phys. Rev. Res.},
  numpages = {8},
  pages = {L042011},
  publisher = {American Physical Society},
  title = {Quantum geometric bounds in spinful systems with trivial band topology},
  url = {https://link.aps.org/doi/10.1103/zlxq-fxgc},
  volume = {7},
  year = {2025}
}

@article{Jankowski2025PRL,
  title = {Enhancing the Hyperpolarizability of Crystals with Quantum Geometry},
  author = {Jankowski, Wojciech J. and Slager, Robert-Jan and Pizzochero, Michele},
  journal = {Phys. Rev. Lett.},
  volume = {135},
  issue = {12},
  pages = {126606},
  numpages = {7},
  year = {2025},
  month = {Sep},
  publisher = {American Physical Society},
  doi = {10.1103/z7lp-pqp6},
  url = {https://link.aps.org/doi/10.1103/z7lp-pqp6}
}

@misc{Jankowski2025gerbe,
      title={Probing Tensor Monopoles and Gerbe Invariants in Three-Dimensional Topological Matter}, 
      author={Wojciech J. Jankowski and Robert-Jan Slager and Giandomenico Palumbo},
      year={2025},
      eprint={2507.22116},
      archivePrefix={arXiv},
      primaryClass={cond-mat.mes-hall},
      url={https://arxiv.org/abs/2507.22116}, 
}

@article{Bzdusek2026,
  title = {{From quantum geometry to nonlinear optics and gerbes: Recent advances in topological band theory}},
  author = {Bzdu\ifmmode \check{s}\else \v{s}\fi{}ek, T.},
  journal = {Phys. Rev. B},
  volume = {113},
  issue = {9},
  pages = {099601},
  numpages = {13},
  year = {2026},
  month = {},
  publisher = {American Physical Society},
  doi = {},
  url = {https://link.aps.org/doi/10.1103/9p92-qt26}
}

@article{Mitscherling2025,
  title = {Gauge-invariant projector calculus for quantum state geometry and applications to observables in crystals},
  author = {Mitscherling, Johannes and Avdoshkin, Alexander and Moore, Joel E.},
  journal = {Phys. Rev. B},
  volume = {112},
  issue = {8},
  pages = {085104},
  numpages = {16},
  year = {2025},
  month = {Aug},
  publisher = {American Physical Society},
  doi = {10.1103/qscv-qxqt},
  url = {https://link.aps.org/doi/10.1103/qscv-qxqt}
}

@article{Yu2025QG,
  author  = {Yu, Jiabin and
             Bernevig, B. Andrei and
             Queiroz, Raquel and
             Rossi, Enrico and
             T{\"o}rm{\"a}, P{\"a}ivi and
             Yang, Bohm-Jung},
  title   = {Quantum geometry in quantum materials},
  journal = {npj Quantum Materials},
  volume  = {10},
  number  = {1},
  pages   = {101},
  year    = {2025},
  doi     = {10.1038/s41535-025-00801-3}
}

@misc{Bouhon2023,
      title={Quantum geometry beyond projective single bands}, 
      author={Adrien Bouhon and Abigail Timmel and Robert-Jan Slager},
      year={2023},
      eprint={2303.02180},
      archivePrefix={arXiv},
      primaryClass={cond-mat.mes-hall},
      url={https://arxiv.org/abs/2303.02180}, 
}

@article{Verma2026,
  author  = {Verma, Nishchhal and Moll, Philip J. W. and Holder, Tobias and Queiroz, Raquel},
  title   = {Quantum geometry and the hidden scales in materials},
  journal = {Nat. Rev. Phys.},
  volume  = {8},
  pages   = {226--239},
  year    = {2026},
  doi     = {10.1038/s42254-026-00923-y}
}

@article{Hu2026,
  title = {Ferromagnetism versus Antiferromagnetism in Narrow-Band Systems: Competition between Quantum Geometry and Band Dispersion},
  author = {Hu, Haoyu and Vafek, Oskar and Haule, Kristjan and Bernevig, B. Andrei},
  journal = {Phys. Rev. Lett.},
  volume = {136},
  issue = {25},
  pages = {256505},
  numpages = {11},
  year = {2026},
  month = {Jun},
  publisher = {American Physical Society},
  doi = {10.1103/zdyq-3m9x},
  url = {https://link.aps.org/doi/10.1103/zdyq-3m9x}
}

@article{Piechon2016,
  title = {{Geometric orbital susceptibility: Quantum metric without Berry curvature}},
  author = {Pi\'echon, Fr\'ed\'eric and Raoux, Arnaud and Fuchs, Jean-No\"el and Montambaux, Gilles},
  journal = {Phys. Rev. B},
  volume = {94},
  issue = {13},
  pages = {134423},
  numpages = {11},
  year = {2016},
  month = {Oct},
  publisher = {American Physical Society},
  doi = {10.1103/PhysRevB.94.134423},
  url = {https://link.aps.org/doi/10.1103/PhysRevB.94.134423}
}

@article{Chau2026,
  title = {Orbital magnetization reveals multiband topology},
  author = {Chau, Chun Wang and Slager, Robert-Jan and Jankowski, Wojciech J.},
  journal = {Phys. Rev. B},
  volume = {113},
  issue = {23},
  pages = {235154},
  numpages = {24},
  year = {2026},
  month = {Jun},
  publisher = {American Physical Society},
  doi = {10.1103/kmb6-pygd},
  url = {https://link.aps.org/doi/10.1103/kmb6-pygd}
}

@Article{Thompson2025exciton,
author={Thompson, J. J. P.
and Jankowski, W. J.
and Slager, R.-J.
and Monserrat, B.},
title={Topologically enhanced exciton transport},
journal={Nat. Commun.},
year={2025},
month={},
day={13},
volume={16},
number={1},
pages={11448},
issn={2041-1723},
doi={},
url={https://doi.org/10.1038/s41467-025-66276-9}
}

@article{Jankowski2024PRL,
  author = {Jankowski, W. J. and Slager, R.-J.},
  issue = {18},
  journal = {Phys. Rev. Lett.},
  numpages = {8},
  pages = {186601},
  publisher = {American Physical Society},
  title = {Quantized Integrated Shift Effect in Multigap Topological Phases},
  url = {https://link.aps.org/doi/10.1103/PhysRevLett.133.186601},
  volume = {133},
  year = {2024}
}

\newpage

\end{document}


\title{{\textsc{supplemental material}} \\ Multistate Torsion in \textit{Zitterbewegung}}

\newcommand{\TCM}{{Theory of Condensed Matter Group, Cavendish Laboratory, University of Cambridge, J.\,J.\,Thomson Avenue, Cambridge CB3 0HE, UK}}


\author{Wojciech J. Jankowski}
\email{wjj25@cam.ac.uk}
\affiliation{\TCM}

\date{\today}

\maketitle

\begin{widetext}

\tableofcontents 

\newpage

\section{Details on the multistate quantum geometry}\label{app::A}

We begin with a set of definitions. $f_{mn} \equiv f_m - f_n$ denotes band occupation factor differences, $\om_{mn} = (\varepsilon_{m}- \varepsilon_{n})/\hbar$ are the frequencies given by the band energy differences. In this work, we consider nondegenerate bands, i.e., $\varepsilon_{m} \neq \varepsilon_{n}$, for ${m \neq n}$. We~introduce momentum derivatives, $\partial_{i} \equiv \partial_{k_i}$, differences of intraband group velocities $\Delta^{nm}_i = \partial_i \om_{nm}$, and non-Abelian Berry connections, $A^{nm}_{i} = \text{i} \bra{u_{n\textbf{k}}} \ket{\partial_i u_{m\textbf{k}}}$, of Bloch states, $\ket{\psi_{n\textbf{k}}} = e^{\text{i} \textbf{k} \cdot \textbf{r}}\ket{u_{n\textbf{k}}}$, where $\textbf{r}$ is the position operator with components $\hat{r}_i$. The~interband velocities $v^{nm}_i$ ($n \neq m$) are given by the non-Abelian Berry-connection elements,
\beq{}
    v^{nm}_i = \bra{u_{n\textbf{k}}} \hat{v}_i \ket{u_{m\textbf{k}}} = \bra{u_{n\textbf{k}}} \frac{\text{i}}{\hbar} [H_\textbf{k}, \hat{r}_i] \ket{u_{m\textbf{k}}} = \text{i} \om_{nm} A^{nm}_i,
\eeq
as the position operator matrix elements read: $\bra{\psi_{n\textbf{k}}} \hat{r}_i \ket{\psi_{m\textbf{k}'}} = -\text{i} \delta_{nm} \partial_{i} \delta(\textbf{k}-\textbf{k}') + A^{nm}_i$~\cite{Ahn2021}. We define covariant derivatives as,
\beq{}
    \mathcal{D}^{mn}_i \equiv \partial_i - \text{i}(A^{mm}_i - A^{nn}_i).
\eeq
The two-state quantum geometric tensors (QGTs) read~\cite{Ahn2021},
\beq{}
    Q^{nm}_{ij} = A^{nm}_i A^{mn}_j = g^{nm}_{ij} - \frac{\text{i}}{2} F^{nm}_{ij},
\eeq
with a two-state quantum metric $g^{nm}_{ij}$ and two-state symplectic form $F^{nm}_{ij}$ representing a band-projected contribution to a Berry curvature~\cite{Ahn2021}. The Hermitian connections can be defined as~\cite{Ahn2021},
\beq{}
    C^{nm}_{ijk} = A^{nm}_i \mathcal{D}^{mn}_i A^{mn}_j,
\eeq
which culminates in a two-state torsion tensor $T^{nm}_{ijk}$~\cite{Ahn2021, Jankowski2024PRL},
\beq{}
    T^{nm}_{ijk} = C^{nm}_{ijk} - C^{nm}_{ikj} = \text{i} A^{nm}_i \sum_{p \neq n,m} (A^{mp}_j A^{pn}_k - A^{mp}_k A^{pn}_j) = \sum_{p \neq n,m} T^{nmp}_{ijk}.
\eeq
$T^{nmp}_{ijk}$ is a multistate torsion tensor component central to this work,
\beq{}
   T^{nmp}_{ijk} = \text{i} A^{nm}_i (A^{mp}_j A^{pn}_k - A^{mp}_k A^{pn}_j) = \text{i} (Q^{nmp}_{ijk} - Q^{nmp}_{ikj}).
\eeq
In the above, we further defined a three-state QGT~\cite{Mehraeen2025},
\beq{}
    Q^{nmp}_{ijk} = A^{nm}_i A^{mp}_j A^{pn}_k.
\eeq
It should be noted that the three-state torsion tensor component $T^{nmp}_{ijk}$ and the three-state QGTs, $Q^{nmp}_{ijk}$, are gauge-invariant under the $U(1)$ gauge transformations of~the individual nondegenerate Bloch bands. Hence, these tensors constitute measurable multistate quantum geometric quantities, as demonstrated in the main text. 

\section{Derivation of the first-order field-induced Zitterbewegung velocity}\label{app::B}

To derive the field-induced first-order \textit{Zitterbewegung} velocity, we begin with the quantum Liouville equation for a single-particle density matrix $\rho(t)$,
\beq{}
    \partial_t \rho(t) + \frac{\text{i}}{\hbar} [H(t), \rho(t)] = 0.
\eeq
The time-dependent Hamiltonian, $H(t) = H_\textbf{k} + \Delta H(t)$, is given by an unperturbed free-fermion Bloch Hamiltonian $H_\textbf{k}$ under a~minimal coupling to an electric field perturbation in length gauge, $\Delta H(t) = e \textbf{r} \cdot \textbf{E}(t)$. $\textbf{r}$ denotes a single-particle position operator, $\textbf{E}(t)$ represents an electric field vector. We expand the density matrix perturbatively to $N$-th order in the electric field powers as: $\rho(t) = \rho^{(0)}(t) + \rho^{(1)}(t) + \rho^{(2)}(t) + \ldots \rho^{(N)}(t)$, which yields a hierarchy of equations~\cite{Mandal2024},
\beq{}
    \partial_t \rho^{(N)}(t) + \frac{\text{i}}{\hbar} [H_\textbf{k}, \rho^{(N)}(t)] + \frac{\rho^{(N)}(t)}{\tau^{(N)}} = \frac{i \textbf{E}(t)}{\hbar} \cdot [\textbf{r}, \rho^{(N-1)}].
\eeq
$\tau^{(N)}$ are scattering times which we send to $\tau^{(N)} \rightarrow \infty$ in the clean crystal limit central to this work. In the band basis, ${\rho^{(0)}_{nm}(t) = \bra{u_{n\textbf{k}}} \rho^{(0)}(t)\ket{u_{m\textbf{k}}}}$, with Bloch states, $\ket{\psi_{n\textbf{k}}} = e^{\text{i} \textbf{k} \cdot \textbf{r}}\ket{u_{n\textbf{k}}}$, satisfying the unperturbed Schr\"odinger equation, ${H_\textbf{k} \ket{u_{n\textbf{k}}} = \varepsilon_n \ket{u_{n\textbf{k}}}}$. The zero-field density matrix, $\rho^{(0)}_{nm}(t) = \delta_{nm} f_m$, is given by the Fermi--Dirac distribution $f_m$. We focus on the field-induced first-order and second-order corrections: $\rho^{(1)}_{nm}(t) \equiv \rho^{nm}_{E}(t)$ and $\rho^{(2)}_{nm}(t) \equiv \rho^{nm}_{E^2}(t)$ in what follows. 

We choose an instantaneous electric field of monochromatic light, $E_j(t) = \Theta(t)\text{Re}[E_j(\om) e^{\text{i}\om t}]$, where $\Theta(t)$ is a Heaviside step function. Accordingly, $\rho^{(1)}_{nm}(0) = 0$. $\rho^{(1)}_{nm}(t)$ for $t > 0$ follows from the quantum Liouville equation as,
\beq{}
    \partial_t \rho^{mn}_E(t) + \text{i} \om_{nm} \rho^{nm}_E(t) = \frac{\text{i} e}{\hbar} A^{nm}_j f_{mn} E_j (t).
\eeq
The solution to the Liouville equation reads,
\beq{}
    \rho^{mn}_E(t) = \frac{e}{2} A^{nm}_j f_{mn} \Bigg[ E_j \frac{e^{-\text{i}\om t} - e^{-\text{i}\om_{mn} t}}{\varepsilon_{m}-\varepsilon_n-\hbar \om} + E^*_j \frac{e^{\text{i}\om t} - e^{-\text{i}\om_{mn} t}}{\varepsilon_{m}-\varepsilon_n+\hbar \om} \Bigg].
\eeq
Combining the time-dependent field-induced coherence $\rho^{mn}_E(t)$ with the velocity operator as $v^{nm}_i\rho^{mn}_E(t) \rightarrow \xi^{(1),m}_i (t) f_m$, and isolating the oscillatory time-dependent term, gives the first-order field-induced \textit{Zitterbewegung} velocity~\cite{Culcer2026},
\beq{}
    \xi^{(1),m}_i (t) = \frac{\text{i} e E_j (\om)}{\hbar} \sum_{n \neq m} Q^{mn}_{ij} \frac{\om_{mn}}{\om_{mn}-\om} e^{-\text{i} \om_{mn} t} + \text{h.c.}.
\eeq
We note that the dc limit $\om \rightarrow 0$ restores the static field result of first-order field-induced \textit{Zitterbewegung} velocity of Ref.~\cite{Culcer2026}.

Furthermore, it is interesting to consider the field-induced first-order \textit{Zitterbewegung} under a circularly polarized drive. The corresponding monochromatic field with amplitude $|E_0|$ reads: $\textbf{E}(t) = \Theta(t) \frac{E_0}{2} \sum_{\eta = +,-} \boldsymbol{\epsilon}^{(\eta)} e^{\text{i} \eta\om t} $ with $\boldsymbol{\epsilon}^{(\pm)} = \frac{1}{\sqrt{2}}(1,\pm\text{i},0)$ representing the left- and right- circular light polarization vectors. We consider light polarized circularly  in the $x-y$ plane. For convenience, we define,
\beq{}
    A^{mn}_\pm = A^{mn}_x + \text{i} A^{mn}_y. 
\eeq
Correspondingly, solving the Liouville equation to first order in electric field yields,
\beq{}
    \rho^{mn}_E(t) = \frac{e E_0}{2} f_{mn} \Bigg[ A^{mn}_+ \frac{e^{-\text{i}\om t} - e^{-\text{i}\om_{mn} t}}{\varepsilon_{m}-\varepsilon_n-\hbar \om} + A^{mn}_- \frac{e^{\text{i}\om t} - e^{-\text{i}\om_{mn} t}}{\varepsilon_{m}-\varepsilon_n+\hbar \om} \Bigg],
\eeq
for left circularly~polarized electric field, $\textbf{E}(t) =  \Theta (t) E_0[\hat{x}~\text{cos}(\om t) + \hat{y}~\text{sin}(\om t)]$, with $\textbf{E}(\om) = \frac{E_0}{2} (1, \text{i},0)$. The resultant circular field-induced \textit{Zitterbewegung} amounts to,
\beq{}
    \xi^{(1),m}_i (t) = -\frac{ e E_0 }{\hbar} \sum_{n \neq m} \om_{mn} \text{Im} \Bigg[ \Bigg( \frac{Q^{mn}_{i+}}{\om_{mn}-\om} + \frac{Q^{mn}_{i-}}{\om_{mn}+\om} e^{-\text{i} \om_{mn} t} \Bigg) \Bigg],
\eeq
where we defined circularly-projected two-state QGTs, $Q^{mn}_{i\pm} = A^{mn}_i A^{nm}_\pm$. Finally, it is interesting to integrate the first-order field-induced \textit{Zitterbewegung} velocity, $\Delta r^m_i (t) \equiv \int^t_0 \text{d}t'~\xi^{(1),m}_i (t')$, in the $\om \neq 0$ regime, to compare with the field-induced positional shift~\cite{Gao2014} arising from the dc electric field limit of the field-induced \textit{Zitterbewegung} velocity. For the circularly~polarized drive, the integration of the electric field amounts to an impulse $\mathcal{I} (t) \equiv \int^t_0 \text{d}t'~\textbf{E}(t')$, and we obtain,
\beq{}
    \Delta r^m_i (t) = 2 \text{Re}~[\alpha^m_{ij}(\om) E_j (\om) e^{-\text{i} \om t}] - \frac{e \mathcal{I}_j (t)}{\hbar} \sum_{n \neq m} f_{mn} \Om^{mn}_{ij} - 2 \text{Re}~\Bigg[ \frac{e}{\hbar} \sum_{n \neq m} f_{mn} Q^{mn}_{ij} \Bigg( \frac{E_j(\om)}{\om_{mn} - \om} +  \frac{E^*_j(\om)}{\om_{mn} + \om}\Bigg) e^{-\text{i} \om_{mn} t} \Bigg]. 
\eeq
The first term is an ac generalization of the field-induced positional shift of a Bloch state~\cite{Gao2014}, 
\beq{}
    \alpha^m_{ij} (\om) = \frac{e}{\hbar} \sum_{n\neq m} f_{mn} \Bigg[ \frac{Q^{mn}_{ij}}{\om_{mn}-\om} + \frac{Q^{nm}_{ij}}{\om_{mn}+\om} \Bigg] = \frac{e}{\hbar} \sum_{n\neq m} f_{mn} \frac{2\om_{mn} g^{nm}_{ij}- \text{i} \om \Omega^{mn}_{ij}}{\om_{mn}^2-\om^2},
\eeq
which corresponds to the static field-induced position shift, $\alpha^m_{ij} (0) = \frac{2e}{\hbar} \frac{g^{nm}_{ij}}{\om_{mn}}$, in the dc limit, $\om \rightarrow 0$~\cite{Gao2014, Culcer2026}.

In the next section, we derive the second-order field-induced \textit{Zitterbewegung} velocity, upon extending the above developments derived from the quantum Liouville equation.

\newpage

\section{Derivation of the second-order field-induced Zitterbewegung velocity}\label{app::C}

Below, we derive the second-order field-induced \textit{Zitterbewegung} velocity. The first-order field-induced interband coherences in the density matrix,  $\rho^{(1)}_{nm}(t) \equiv \rho^{nm}_E(t)$, under circularly~polarized monochromatic drive were considered in the previous section. We now focus on the second-order field-induced interband coherences in the density matrix,  $\rho^{(2)}_{nm}(t) \equiv \rho^{nm}_{E^2}$, induced by the circularly~polarized monochromatic drive with frequency $\omega$.

We again choose a monochromatic field with amplitude $|E_0|$ and electric field vector, $\textbf{E}(t) = \Theta(t) \frac{E_0}{2} \sum_{\eta = +,-} \boldsymbol{\epsilon}^{(\eta)} e^{\text{i} \eta\om t}$, with $\boldsymbol{\epsilon}^{(\pm)} = \frac{1}{\sqrt{2}}(1,\pm\text{i},0)$ representing the left- and right- circular light polarization vectors. The quantum Liouville equation at the second perturbative order reads,
\beq{}
    \partial_t \rho^{mn}_{E^2}(t) + \text{i} \om_{nm} \rho^{mn}_{E^2}(t) = \frac{e E_i (t)}{\hbar} [\mathcal{D}_i \rho_{E}(t)]^{mn},
\eeq
with the covariant derivative acting on the density matrix as~\cite{Mandal2024},
\beq{}
    [\mathcal{D}_i \rho]^{mn} = \partial_i \rho^{mn}- \text{i} [ \mathcal{A}_i, \rho]^{mn}.
\eeq
$\mathcal{A}_i = \sum_{n,m} A^{nm}_i \ket{u_{n\textbf{k}}} \bra{u_{m\textbf{k}}}$ is the Berry connection operator. Correspondingly, we employ the results for the first-order density matrix interband coherences $\rho^{nm}_{E}(t)$ under the circular drive, as derived in the previous section. The first-order density matrix corrections in the clean limit ($\tau \rightarrow \infty$) read,
\beq{}
    \rho^{mm}_E (t) = \frac{e E_0}{2\hbar} \sum_\eta \text{i} (\boldsymbol{\epsilon}^{(\eta)} \cdot \partial_\textbf{k}) f_{nm} \frac{1 - e^{-\text{i} \eta \om t}}{\text{i}\eta \om},
\eeq
\beq{}
    \rho^{nm}_E (t) = \frac{e E_0}{2\hbar} \sum_\eta \text{i} (\boldsymbol{\epsilon}^{(\eta)} \cdot \textbf{A}^{nm}) f_{nm} \frac{e^{-\text{i} \eta \om t} - e^{-\text{i} \om_{nm} t}}{\text{i}(\om_{nm} - \eta \om)},
\eeq
with the Berry connection vector $\textbf{A}^{nm} = (A^{nm}_x, A^{nm}_y, A^{nm}_z)$.
Under the circular polarizations $\eta, \eta'$, we combine electric fields as, $E_c E_d \sim \frac{E^2_0}{4} \boldsymbol{\epsilon}^{(\eta)}_c \boldsymbol{\epsilon}^{(\eta')}_d$. We first obtain $\rho^{mn}_{E^2}(t)$ upon solving the Liouville equation in the dc limit $(\om \rightarrow 0)$ as,

\beq{}
\begin{aligned}
    \rho^{mn}_{E^2}(t) &= \frac{e^2 E_c E_d}{\hbar^2} \Bigg[ \text{i} (\mathcal{D}^{nm}_c A^{mn}_d) f_{mn} \Bigg( -\frac{1- e^{-\text{i} \om_{mn} t}}{\om_{mn}^2} + \frac{\text{i} t e^{-\text{i} \om_{mn} t}}{ \om_{mn}} \Bigg) 
    +\text{i} A^{mn}_d f_{mn} \Delta^{mn}_c \Bigg( \frac{t^2 e^{-\text{i} \om_{mn} t}}{2\om_{mn}} - \frac{\text{i} t e^{-\text{i} \om_{mn} t}}{\om_{mn}^2} + \frac{1- e^{-\text{i} \om_{mn} t}}{\om_{mn}^3}\Bigg)
    \\&+ \text{i} A^{mn}_d \partial_c f_{mn} \Bigg( -\frac{1- e^{-\text{i} \om_{mn} t}}{\om_{mn}^2} + \frac{\text{i} t e^{-\text{i} \om_{mn} t}}{ \om_{mn}} \Bigg)
    +\text{i} A^{mn}_c \partial_d f_{mn} \Bigg(-\frac{\text{i} t}{\om_{mn}} +\frac{1 - e^{-\text{i} \om_{mn} t}}{\om_{mn}^2} \Bigg)
    + \rho^{mn}_{E^2,3b}(t) \Bigg],
\end{aligned}
\eeq
with the multistate three-band term $\rho^{mn}_{E^2,3b}(t)$, 
\beq{}
    \rho^{mn}_{E^2,3b}(t) = \frac{e^2 E_c E_d}{\hbar^2} \sum_{p \neq n,m} \Big[ A^{mp}_c A^{pn}_d f_{pn} L_{mn,pn}(t) - A^{mp}_d A^{pn}_c f_{mp} L_{nm,np}(t) \Big].
\eeq
The three-state coherence term includes a time-dependent kernel $L_{mn,pn}(t)$,
\beq{}
     L_{mn,pn}(t) = -\frac{1- e^{-\text{i} \om_{mn} t}}{\om_{pn} \om_{mn}} + \frac{e^{- \text{i} \om_{pn} t}-e^{-\text{i} \om_{mn} t}}{\om_{pn} \om_{mp}}.
\eeq
We note that this multiband coherence term is consistent with the multistate $\rho^{(2)}_{nm}$ derived in Ref.~\cite{Mandal2024}. Accordingly, the multistate contribution to the \textit{Zitterbewegung} velocity in the static field limit reads,
\beq{}
    \xi^{(2),m}_{3b,i}(t) = \frac{e^2 E_c E_d}{\hbar^2} \sum_{n \neq p \neq m} \Big[-\text{i} \om_{ln} Q^{npm}_{icd} L_{pn,mn}(t) + \text{i} \om_{nm} Q^{nmp}_{icd} L_{nm,pm}(t) -  \text{i} \om_{nm} Q^{npm}_{idc} L_{pn,pm}(t) +  \text{i} \om_{mn} Q^{nmp}_{idc} L_{mn,mp}(t)\Big].
\eeq
We now extend these results to finite frequencies $(\om \neq 0)$ for photons with circular polarizations $\eta, \eta'$. The dynamically-induced second-order \textit{Zitterbewegung} velocity is derived analogously upon solving the quantum Liouville equations. To state the result compactly, we define a set of frequency-based factors,

\begin{align}
L^{\eta'\eta}_{mn,pn}(t) &= -\frac{1}{\om_{pn}-\eta\om}\left[\frac{e^{-\text{i}(\eta+\eta')\om t}-e^{-\text{i}\om_{mn} t}}{\om_{mn}-(\eta+\eta')\om}-\frac{e^{-\text{i}(\om_{pn}+\eta'\om)t}-e^{-\text{i}\om_{mn}t}}{\om_{mn}-\om_{pn}-\eta'\om}\right],\\
K^{\eta'\eta}_{mn}(t) &= -\frac{1}{\om_{mn}-\eta\om}\left[\frac{e^{-\text{i}(\eta+\eta')\om t}-e^{-\text{i}\om_{mn} t}}{\om_{mn}-(\eta+\eta')\om}-e^{-\text{i}\om_{mn} t}\frac{1-e^{-\text{i}\eta'\om t}}{\eta'\om}\right],\\
J^{\eta'\eta}_{mn}(t) &= \frac{e^{-\text{i}\om_{mn} t}}{\om_{mn}-\eta\om}\left[\frac{it\,e^{-\text{i}\eta'\om t}}{\eta'\om}-\frac{1-e^{-\text{i}\eta'\om t}}{\om^2}\right]-\frac{K^{\eta'\eta}_{mn}}{\om_{mn}-\eta\om},\\
M^{\eta'\eta}_{mn}(t) &= \frac{1}{\om_{mn}-\eta\om}\left[\frac{e^{-\text{i}(\eta+\eta')\om t}-1}{(\eta+\eta')\om}-\frac{e^{-\text{i}(\om_{mn}+\eta'\om)t}-1}{\om_{mn}+\eta'\om}\right],\\
N^{\eta'\eta}_{mn}(t) &= \frac{1}{\text{i}\eta\om}\left[\frac{e^{-\text{i}\eta'\om t}-e^{-\text{i}\om_{mn} t}}{\text{i}(\om_{mn}-\eta'\om)}-\frac{e^{-\text{i}(\eta+\eta')\om t}-e^{-\text{i}\om_{mn} t}}{\text{i}(\om_{mn}-(\eta+\eta')\om)}\right].
\end{align}
Correspondingly, the dynamically-induced second-order \textit{Zitterbewegung} velocity is given as, 
\beq{}
\begin{aligned}
\xi^{(2),m}_i(t)&=\frac{e^2 E_c E_d}{4\hbar^2}\sum_{\eta,\eta'} \sum_{n \neq m}\Big[\om_{mn} C^{mn}_{icd} K^{\eta'\eta}_{mn}(t)+\om_{mn} Q^{mn}_{id}\Delta^c_{mn}J^{\eta'\eta}_{mn}(t)-(m\!\leftrightarrow\! n)\Big]+\sum_{n\ne p}v^i_m\Big[Q^{mn}_{cd}M^{\eta'\eta}_{mn}(t)-(m\!\leftrightarrow\! n)\Big]\\
&+\sum_{n \neq p \neq m}\Big[-\text{i}\om_{pn} Q^{npm}_{icd} L^{\eta'\eta}_{pn,mn}(t)+\text{i}\om_{nm} Q^{mnp}_{icd} L^{\eta'\eta}_{nm,pm}(t) -\text{i}\om_{pn} Q^{npm}_{idc} L^{\eta'\eta}_{pn,pm}(t)+\text{i}\om_{mn} Q^{nmp}_{idc} L^{\eta'\eta}_{mn,mp}(t)\Big] + \ldots,
\end{aligned}
\eeq
where $\ldots$ includes additional terms derived upon integrating by parts distribution function nonuniformities, $\partial_k f_{mn}$. The finite-frequency multistate \textit{Zitterbewegung} term reads,
\beq{}\label{eq::FFMZ}
    \xi^{(2),m}_{3b,i}(t) = \frac{e^2 E_c E_d}{\hbar^2} \sum_{n \neq p \neq m}\Big[-\text{i}\om_{pn} Q^{npm}_{icd} L^{\eta'\eta}_{pn,mn}(t)+\text{i}\om_{nm} Q^{mnp}_{icd} L^{\eta'\eta}_{nm,pm}(t) -\text{i}\om_{pn} Q^{npm}_{idc} L^{\eta'\eta}_{pn,pm}(t)+\text{i}\om_{mn} Q^{nmp}_{idc} L^{\eta'\eta}_{mn,mp}(t)\Big].
\eeq
We will now focus on the contribution at the frequency $\om_{pn}$ corresponding to the energy difference of virtual levels,
\beq{}
    \xi^{(2),m}_i(t)\Big|_{\om_{pn}} \propto \frac{e^2 E^2_0}{\hbar^2} \mathcal{A}^{mnp}_i(\om) e^{-\text{i} \om_{pn} t}.
\eeq
The frequency-dependent amplitude factor consistent with Eq.~\eqref{eq::FFMZ} reads,
\beq{}
    \mathcal{A}^{mnp}_i(\om) =  \frac{\text{i}}{4} A^{np}_i \sum_{\eta, \eta'} \frac{\om_{pn}}{\om_{pn}-(\eta + \eta')\om} \Bigg[ \frac{(\boldsymbol{\epsilon}^{(\eta')} \cdot \textbf{A}^{pm})(\boldsymbol{\epsilon}^{(\eta)} \cdot \textbf{A}^{mn})}{\om_{pm} - \eta' \om} + \frac{(\boldsymbol{\epsilon}^{(\eta)} \cdot \textbf{A}^{pm})(\boldsymbol{\epsilon}^{(\eta')} \cdot \textbf{A}^{mn})}{\om_{mn} - \eta' \om} \Bigg].
\eeq
Correspondingly, we can identify \textit{Zitterbewegung} velocities $ \xi^{(2),m}_i(t)\Big|^{\eta \eta'}_{\om_{pn}}$ arising in the individual helicity channels $(\eta, \eta' = \pm)$,
\beq{}
 \xi^{(2),m}_i(t)\Big|^{+-}_{\om_{pn}} = \frac{e^2 E^2_0}{\hbar^2} w^{+-}_{mnp}(\om) \Bigg[ \text{Im}~T^{npm}_{ixy} \cos (\om_{pn} t) + \text{Re}~T^{npm}_{ixy} \sin (\om_{pn} t) \Bigg], 
\eeq
for the counter-rotating amplitude, $\xi^{(2),m}_i(t) \propto  \mathcal{A}^{mnp}_{ijk}(\om,-\om) E_j(\om) E_k(-\om)$, using the main text notation. The frequency-dependent factor reads,
\beq{}
    w_{mnp}^{+-}(\om) = \frac{\om}{\om^2_{mn} - \om^2} - \frac{\om}{\om^2_{pm} - \om^2}.
\eeq
The co-rotating helicity channel yields an amplitude $\mathcal{A}^{mnp}_{ijk}(\om,\om)$, such that $\xi^{(2),m}_i(t) \propto  \mathcal{A}^{mnp}_{ijk}(\om,\om) E_j(\om) E_k(\om)$, which culminates in a \textit{Zitterbewegung} velocity given by,
\beq{}
 \xi^{(2),m}_i(t)\Big|^{++}_{\om_{pn}} = \frac{e^2 E^2_0}{\hbar^2} \text{Re} \Bigg[w^{++}_{nmp} (\om)  \Big(Q^{npm}_{ixy} + Q^{npm}_{ixy}\Big) e^{-\text{i} \om_{pn} t}   \Bigg], 
\eeq
with another frequency-dependent factor,
\beq{}
    w_{mnp}^{++}(\om) = \frac{1}{\om_{pn} - 2\om} \Bigg(\frac{1}{\om_{pm} - \om} - \frac{1}{\om_{mn} - \om} \Bigg).
\eeq

Interestingly, on choosing a linear polarization of photons instead, $\boldsymbol{\epsilon} = \frac{1}{\sqrt{2}} (\boldsymbol{\epsilon}^{(+)} + \boldsymbol{\epsilon}^{(-)}) = \hat{x}$, the dynamically-induced nonlinear \textit{Zitterbewegung} velocity component arises from three-state QGTs $Q^{nmp}_{ixx}$ as,
\beq{}
    \xi^{(2),m}_i(t)\Big|^{\hat{x} \hat{x}}_{\om_{pn}} = -\frac{e^2 E^2_0}{\hbar^2} \sum_{n,p} w^{\hat{x} \hat{x}}_{mnp}(\om) \Bigg[ \text{Im}~Q^{nmp}_{ixx} \cos(\om_{pn} t) - \text{Re}~Q^{nmp}_{ixx} \sin(\om_{pn} t)  \Bigg].
\eeq
The multiband frequency factor $w^{\hat{x} \hat{x}}_{mnp}(\om)$ reads,
\beq{}
    w^{\hat{x} \hat{x}}_{mnp}(\om) = \sum_{\eta = \pm} \frac{\om_{pn} - \eta \om}{(\om_{pm} - \eta \om)(\om_{mn} - \eta \om)}.
\eeq
Therefore, the nonlinear \textit{Zitterbewegung} velocities that arise at second order in electric field coupling provide a window to measure both the multistate torsion tensors $T^{nmp}_{ijk}$ and the three-state QGTs $Q^{nmp}_{ijj}$ under the optical drives with distinct light polarizations.

\end{widetext}

\bibliographystyle{apsrev4-1}
\bibliography{references.bib}